
\font \tbfontt                = cmbx10 scaled\magstep1
\font \tafontt                = cmbx10 scaled\magstep2
\font \tbfontss               = cmbx5  scaled\magstep1
\font \tafontss               = cmbx5  scaled\magstep2
\font \sixbf                  = cmbx6
\font \tbfonts                = cmbx7  scaled\magstep1
\font \tafonts                = cmbx7  scaled\magstep2
\font \ninebf                 = cmbx9
\font \tasys                  = cmex10 scaled\magstep1

\font \sixi                   = cmmi6
\font \ninei                  = cmmi9
\font \tams                   = cmmib10
\font \tbmss                  = cmmib10 scaled 600
\font \tamss                  = cmmib10 scaled 700
\font \tbms                   = cmmib10 scaled 833
\font \tbmt                   = cmmib10 scaled\magstep1
\font \tamt                   = cmmib10 scaled\magstep2
\font \smallescriptscriptfont = cmr5
\font \smalletextfont         = cmr5 at 10pt
\font \smallescriptfont       = cmr5 at 7pt
\font \sixrm                  = cmr6
\font \ninerm                 = cmr9
\font \ninesl                 = cmsl9
\font \tensans                = cmss10
\font \fivesans               = cmss10 at 5pt
\font \sixsans                = cmss10 at 6pt
\font \sevensans              = cmss10 at 7pt
\font \ninesans               = cmss10 at 9pt
\font \tbst                   = cmsy10 scaled\magstep1
\font \tast                   = cmsy10 scaled\magstep2
\font \tbsss                  = cmsy5  scaled\magstep1
\font \tasss                  = cmsy5  scaled\magstep2
\font \sixsy                  = cmsy6
\font \tbss                   = cmsy7  scaled\magstep1
\font \tass                   = cmsy7  scaled\magstep2
\font \ninesy                 = cmsy9
\font \markfont               = cmti10 at 11pt
\font \nineit                 = cmti9
\font \ninett                 = cmtt9
\magnification=\magstep0
\hsize=13truecm
\vsize=19.8truecm
\hfuzz=2pt
\tolerance=500
\abovedisplayskip=3 mm plus6pt minus 4pt
\belowdisplayskip=3 mm plus6pt minus 4pt
\abovedisplayshortskip=0mm plus6pt minus 2pt
\belowdisplayshortskip=2 mm plus4pt minus 4pt
\predisplaypenalty=0
\clubpenalty=10000
\widowpenalty=10000
\frenchspacing
\newdimen\oldparindent\oldparindent=1.5em
\parindent=1.5em



\def\bbbc{{\mathchoice {\setbox0=\hbox{$\displaystyle\rm C$}\hbox{\hbox
to0pt{\kern0.4\wd0\vrule height0.9\ht0\hss}\box0}}
{\setbox0=\hbox{$\textstyle\rm C$}\hbox{\hbox
to0pt{\kern0.4\wd0\vrule height0.9\ht0\hss}\box0}}
{\setbox0=\hbox{$\scriptstyle\rm C$}\hbox{\hbox
to0pt{\kern0.4\wd0\vrule height0.9\ht0\hss}\box0}}
{\setbox0=\hbox{$\scriptscriptstyle\rm C$}\hbox{\hbox
to0pt{\kern0.4\wd0\vrule height0.9\ht0\hss}\box0}}}}
\def\bbbe{{\mathchoice {\setbox0=\hbox{\smalletextfont e}\hbox{\raise
0.1\ht0\hbox to0pt{\kern0.4\wd0\vrule width0.3pt
height0.7\ht0\hss}\box0}}
{\setbox0=\hbox{\smalletextfont e}\hbox{\raise 0.1\ht0\hbox
to0pt{\kern0.4\wd0\vrule width0.3pt height0.7\ht0\hss}\box0}}
{\setbox0=\hbox{\smallescriptfont e}\hbox{\raise 0.1\ht0\hbox
to0pt{\kern0.5\wd0\vrule width0.2pt height0.7\ht0\hss}\box0}}
{\setbox0=\hbox{\smallescriptscriptfont e}\hbox{\raise
0.1\ht0\hbox to0pt{\kern0.4\wd0\vrule width0.2pt
height0.7\ht0\hss}\box0}}}}
\def\bbbq{{\mathchoice {\setbox0=\hbox{$\displaystyle\rm Q$}\hbox{\raise
0.15\ht0\hbox to0pt{\kern0.4\wd0\vrule height0.8\ht0\hss}\box0}}
{\setbox0=\hbox{$\textstyle\rm Q$}\hbox{\raise
0.15\ht0\hbox to0pt{\kern0.4\wd0\vrule height0.8\ht0\hss}\box0}}
{\setbox0=\hbox{$\scriptstyle\rm Q$}\hbox{\raise
0.15\ht0\hbox to0pt{\kern0.4\wd0\vrule height0.7\ht0\hss}\box0}}
{\setbox0=\hbox{$\scriptscriptstyle\rm Q$}\hbox{\raise
0.15\ht0\hbox to0pt{\kern0.4\wd0\vrule height0.7\ht0\hss}\box0}}}}
\def\bbbt{{\mathchoice {\setbox0=\hbox{$\displaystyle\rm
T$}\hbox{\hbox to0pt{\kern0.3\wd0\vrule height0.9\ht0\hss}\box0}}
{\setbox0=\hbox{$\textstyle\rm T$}\hbox{\hbox
to0pt{\kern0.3\wd0\vrule height0.9\ht0\hss}\box0}}
{\setbox0=\hbox{$\scriptstyle\rm T$}\hbox{\hbox
to0pt{\kern0.3\wd0\vrule height0.9\ht0\hss}\box0}}
{\setbox0=\hbox{$\scriptscriptstyle\rm T$}\hbox{\hbox
to0pt{\kern0.3\wd0\vrule height0.9\ht0\hss}\box0}}}}
\def\bbbs{{\mathchoice
{\setbox0=\hbox{$\displaystyle     \rm S$}\hbox{\raise0.5\ht0\hbox
to0pt{\kern0.35\wd0\vrule height0.45\ht0\hss}\hbox
to0pt{\kern0.55\wd0\vrule height0.5\ht0\hss}\box0}}
{\setbox0=\hbox{$\textstyle        \rm S$}\hbox{\raise0.5\ht0\hbox
to0pt{\kern0.35\wd0\vrule height0.45\ht0\hss}\hbox
to0pt{\kern0.55\wd0\vrule height0.5\ht0\hss}\box0}}
{\setbox0=\hbox{$\scriptstyle      \rm S$}\hbox{\raise0.5\ht0\hbox
to0pt{\kern0.35\wd0\vrule height0.45\ht0\hss}\raise0.05\ht0\hbox
to0pt{\kern0.5\wd0\vrule height0.45\ht0\hss}\box0}}
{\setbox0=\hbox{$\scriptscriptstyle\rm S$}\hbox{\raise0.5\ht0\hbox
to0pt{\kern0.4\wd0\vrule height0.45\ht0\hss}\raise0.05\ht0\hbox
to0pt{\kern0.55\wd0\vrule height0.45\ht0\hss}\box0}}}}
\def\bbbz{{\mathchoice {\hbox{$\sans\textstyle Z\kern-0.4em Z$}}
{\hbox{$\sans\textstyle Z\kern-0.4em Z$}}
{\hbox{$\sans\scriptstyle Z\kern-0.3em Z$}}
{\hbox{$\sans\scriptscriptstyle Z\kern-0.2em Z$}}}}
\skewchar\ninei='177 \skewchar\sixi='177
\skewchar\ninesy='60 \skewchar\sixsy='60
\hyphenchar\ninett=-1
\def\newline{\hfil\break}%
\catcode`@=11
\def\folio{\ifnum\pageno<\z@
\uppercase\expandafter{\romannumeral-\pageno}%
\else\number\pageno \fi}
\catcode`@=12 
  \mathchardef\Gamma="0100
  \mathchardef\Delta="0101
  \mathchardef\Theta="0102
  \mathchardef\Lambda="0103
  \mathchardef\Xi="0104
  \mathchardef\Pi="0105
  \mathchardef\Sigma="0106
  \mathchardef\Upsilon="0107
  \mathchardef\Phi="0108
  \mathchardef\Psi="0109
  \mathchardef\Omega="010A
\def\squareforqed{\hbox{\rlap{$\sqcap$}$\sqcup$}}
\def\qed{\ifmmode\squareforqed\else{\unskip\nobreak\hfil
\penalty50\hskip1em\null\nobreak\hfil\squareforqed
\parfillskip=0pt\finalhyphendemerits=0\endgraf}\fi}
\newfam\sansfam
\textfont\sansfam=\tensans\scriptfont\sansfam=\sevensans
\scriptscriptfont\sansfam=\fivesans
\def\sans{\fam\sansfam\tensans}
\def\stackfigbox{\if
Y\FIG\global\setbox\figbox=\vbox{\unvbox\figbox\box1}%
\else\global\setbox\figbox=\vbox{\box1}\global\let\FIG=Y\fi}
\def\placefigure{\dimen0=\ht1\advance\dimen0by\dp1
\advance\dimen0by5\baselineskip
\advance\dimen0by0.4true cm
\ifdim\dimen0>\vsize\pageinsert\box1\vfill\endinsert
\else
\if Y\FIG\stackfigbox\else
\dimen0=\pagetotal\ifdim\dimen0<\pagegoal
\advance\dimen0by\ht1\advance\dimen0by\dp1\advance\dimen0by1.7true cm
\ifdim\dimen0>\pagegoal\stackfigbox
\else\box1\vskip7true mm\fi
\else\box1\vskip7true mm\fi\fi\fi\let\firstleg=Y}
%
\def\begfig#1cm#2\endfig{\par
\setbox1=\vbox{\dimen0=#1true cm\advance\dimen0
by1true cm\kern\dimen0\vskip-.8333\baselineskip#2}\placefigure}
\def\begdoublefig#1cm #2 #3 \enddoublefig{\begfig#1cm%
\line{\vtop{\hsize=0.46\hsize#2}\hfill
\vtop{\hsize=0.46\hsize#3}}\endfig}
\let\firstleg=Y
\def\figure#1#2{\if Y\firstleg\vskip1true cm\else\vskip1.7true mm\fi
\let\firstleg=N\setbox0=\vbox{\noindent\petit{\bf
Fig.\ts#1\unskip.\ }\ignorespaces #2\smallskip
\count255=0\global\advance\count255by\prevgraf}%
\ifnum\count255>1\box0\else
\centerline{\petit{\bf Fig.\ts#1\unskip.\
}\ignorespaces#2}\smallskip\fi}

\def\begtab#1cm#2\endtab{\par
   \ifvoid\topins\midinsert\medskip\vbox{#2\kern#1true cm}\endinsert
   \else\topinsert\vbox{#2\kern#1true cm}\endinsert\fi}
\def\begpet{\vskip6pt\bgroup\petit}
\def\endpet{\vskip6pt\egroup}
\newdimen\refindent
\newlinechar=`\^
\def\begref#1#2{\titlea{}{#1}%
\bgroup\petit
\setbox0=\hbox{#2\enspace}\refindent=\wd0\relax
\if>#2>\else
\ifdim\refindent>0.5em\else
\message{^Something may be wrong with your references;}%
\message{probably you missed the second argument of \string\begref.}%
\fi\fi}
\def\ref{\goodbreak
\hangindent\oldparindent\hangafter=1
\noindent\ignorespaces}
\def\refno#1{\goodbreak
\setbox0=\hbox{#1\enspace}\ifdim\refindent<\wd0\relax
\message{^Your reference `#1' is wider than you pretended in using
\string\begref.}\fi
\hangindent\refindent\hangafter=1
\noindent\kern\refindent\llap{#1\enspace}\ignorespaces}
\def\refmark#1{\goodbreak
\setbox0=\hbox{#1\enspace}\ifdim\refindent<\wd0\relax
\message{^Your reference `#1' is wider than you pretended in using
\string\begref.}\fi
\hangindent\refindent\hangafter=1
\noindent\hbox to\refindent{#1\hss}\ignorespaces}
\def\endref{\goodbreak\endpet}

\def\petit{\def\rm{\fam0\ninerm}%
\textfont0=\ninerm \scriptfont0=\sixrm \scriptscriptfont0=\fiverm
 \textfont1=\ninei \scriptfont1=\sixi \scriptscriptfont1=\fivei
 \textfont2=\ninesy \scriptfont2=\sixsy \scriptscriptfont2=\fivesy
 \def\it{\fam\itfam\nineit}%
 \textfont\itfam=\nineit
 \def\sl{\fam\slfam\ninesl}%
 \textfont\slfam=\ninesl
 \def\bf{\fam\bffam\ninebf}%
 \textfont\bffam=\ninebf \scriptfont\bffam=\sixbf
 \scriptscriptfont\bffam=\fivebf
 \def\sans{\fam\sansfam\ninesans}%
 \textfont\sansfam=\ninesans \scriptfont\sansfam=\sixsans
 \scriptscriptfont\sansfam=\fivesans
 \def\tt{\fam\ttfam\ninett}%
 \textfont\ttfam=\ninett
 \normalbaselineskip=11pt
 \setbox\strutbox=\hbox{\vrule height7pt depth2pt width0pt}%
 \normalbaselines\rm
\def\vec##1{{\textfont1=\tbms\scriptfont1=\tbmss
\textfont0=\ninebf\scriptfont0=\sixbf
\mathchoice{\hbox{$\displaystyle##1$}}{\hbox{$\textstyle##1$}}
{\hbox{$\scriptstyle##1$}}{\hbox{$\scriptscriptstyle##1$}}}}}
\nopagenumbers
%
\let\header=Y
\let\FIG=N
\newbox\figbox
\output={\if N\header\headline={\hfil}\fi\plainoutput
\global\let\header=Y\if Y\FIG\topinsert\unvbox\figbox\endinsert
\global\let\FIG=N\fi}
\let\lasttitle=N
\def\centerpar#1{{\parfillskip=0pt
\rightskip=0pt plus 1fil
\leftskip=0pt plus 1fil
\advance\leftskip by\oldparindent
\advance\rightskip by\oldparindent
\def\newline{\break}%
\noindent\ignorespaces#1\par}}
\catcode`\@=\active
\def\author#1{\bgroup
\baselineskip=13.2pt
\lineskip=0pt
\pretolerance=10000
\markfont
\centerpar{#1}\bigskip\egroup
{\def@##1{}%
\setbox0=\hbox{\petit\kern2.5true cc\ignorespaces#1\unskip}%
\ifdim\wd0>\hsize
\message{The names of the authors exceed the headline, please use a }%
\message{short form with AUTHORRUNNING}\gdef\leftheadline{%
\rlap{\folio}\hfil AUTHORS suppressed due to excessive length}%
\else
\xdef\leftheadline{\rlap{\noexpand\folio}\hfil
\ignorespaces#1\unskip}%
\fi
}\let\INS=E}
\def\address#1{\bgroup\petit
\centerpar{#1}\bigskip\egroup
\catcode`\@=12
\vskip2cm\noindent\ignorespaces}
\let\INS=N%
\def@#1{\if N\INS\unskip$\,^{#1}$\else\global\footcount=#1\relax
\if E\INS\hangindent0.5\parindent\noindent\hbox
to0.5\parindent{$^{#1}$\hfil}\let\INS=Y\ignorespaces
\else\par\hangindent0.5\parindent\noindent\hbox
to0.5\parindent{$^{#1}$\hfil}\ignorespaces\fi\fi}%
\catcode`\@=12
\headline={\petit\def\newline{ }\def\fonote#1{}\ifodd\pageno
\rightheadline\else\leftheadline\fi}
\def\rightheadline{Missing CONTRIBUTION
title\hfil\llap{\folio}}
\def\leftheadline{\rlap{\folio}\hfil Missing name(s)
of the author(s)}
\nopagenumbers
\let\header=Y

\let\lasttitle=N
 \def\contribution#1{\vfill\eject
 \let\header=N\bgroup
 \textfont0=\tafontt \scriptfont0=\tafonts \scriptscriptfont0=\tafontss
 \textfont1=\tamt \scriptfont1=\tams \scriptscriptfont1=\tams
 \textfont2=\tast \scriptfont2=\tass \scriptscriptfont2=\tasss
 \par\baselineskip=16pt
     \lineskip=16pt
     \tafontt
     \raggedright
     \pretolerance=10000
     \noindent
     \centerpar{\ignorespaces#1}%
     \vskip17pt\egroup
     \nobreak
     \parindent=0pt
     \everypar={\global\parindent=1.5em
     \global\let\lasttitle=N\global\everypar={}}%
     \global\let\lasttitle=A%
     \setbox0=\hbox{\petit\def\newline{ }\def\fonote##1{}\kern2.5true
     cc\ignorespaces#1}\ifdim\wd0>\hsize
     \message{Your CONTRIBUTIONtitle exceeds the headline,
please use a short form
with CONTRIBUTIONRUNNING}\gdef\rightheadline{CONTRIBUTION title
suppressed due to excessive length\hfil\llap{\folio}}%
\else
\gdef\rightheadline{\ignorespaces#1\unskip\hfil\llap{\folio}}\fi
\catcode`\@=\active
     \ignorespaces}
\def\titlea#1#2{\if N\lasttitle\else\vskip-28pt
     \fi
     \vskip18pt plus 4pt minus4pt
     \bgroup
\textfont0=\tbfontt \scriptfont0=\tbfonts \scriptscriptfont0=\tbfontss
\textfont1=\tbmt \scriptfont1=\tbms \scriptscriptfont1=\tbmss
\textfont2=\tbst \scriptfont2=\tbss \scriptscriptfont2=\tbsss
\textfont3=\tasys \scriptfont3=\tenex \scriptscriptfont3=\tenex
     \baselineskip=16pt
     \lineskip=0pt
     \pretolerance=10000
     \noindent
     \tbfontt
     \rightskip 0pt plus 6em
     \setbox0=\vbox{\vskip23pt\def\fonote##1{}%
     \noindent
     \if>#1>\ignorespaces#2
     \else\ignorespaces#1\unskip\enspace\ignorespaces#2\fi
     \vskip18pt}%
     \dimen0=\pagetotal\advance\dimen0 by-\pageshrink
     \ifdim\dimen0<\pagegoal
     \dimen0=\ht0\advance\dimen0 by\dp0\advance\dimen0 by
     3\normalbaselineskip
     \advance\dimen0 by\pagetotal
     \ifdim\dimen0>\pagegoal\eject\fi\fi
     \noindent
     \if>#1>\ignorespaces#2
     \else\ignorespaces#1\unskip\enspace\ignorespaces#2\fi
     \vskip12pt plus4pt minus4pt\egroup
     \nobreak
     \parindent=0pt
     \everypar={\global\parindent=\oldparindent
     \global\let\lasttitle=N\global\everypar={}}%
     \global\let\lasttitle=A%
     \ignorespaces}
 \def\titleb#1#2{\if N\lasttitle\else\vskip-22pt
     \fi
     \vskip18pt plus 4pt minus4pt
     \bgroup
\textfont0=\tenbf \scriptfont0=\sevenbf \scriptscriptfont0=\fivebf
\textfont1=\tams \scriptfont1=\tamss \scriptscriptfont1=\tbmss
     \lineskip=0pt
     \pretolerance=10000
     \noindent
     \bf
     \rightskip 0pt plus 6em
     \setbox0=\vbox{\vskip23pt\def\fonote##1{}%
     \noindent
     \if>#1>\ignorespaces#2
     \else\ignorespaces#1\unskip\enspace\ignorespaces#2\fi
     \vskip10pt}%
     \dimen0=\pagetotal\advance\dimen0 by-\pageshrink
     \ifdim\dimen0<\pagegoal
     \dimen0=\ht0\advance\dimen0 by\dp0\advance\dimen0 by
     3\normalbaselineskip
     \advance\dimen0 by\pagetotal
     \ifdim\dimen0>\pagegoal\eject\fi\fi
     \noindent
     \if>#1>\ignorespaces#2
     \else\ignorespaces#1\unskip\enspace\ignorespaces#2\fi
     \vskip8pt plus4pt minus4pt\egroup
     \nobreak
     \parindent=0pt
     \everypar={\global\parindent=\oldparindent
     \global\let\lasttitle=N\global\everypar={}}%
     \global\let\lasttitle=B%
     \ignorespaces}
 \def\titlec#1{\if N\lasttitle\else\vskip-\baselineskip
     \fi
     \vskip18pt plus 4pt minus4pt
     \bgroup
\textfont0=\tenbf \scriptfont0=\sevenbf \scriptscriptfont0=\fivebf
\textfont1=\tams \scriptfont1=\tamss \scriptscriptfont1=\tbmss
     \bf
     \noindent
     \ignorespaces#1\unskip\ \egroup
     \ignorespaces}
 \def\titled#1{\if N\lasttitle\else\vskip-\baselineskip
     \fi
     \vskip12pt plus 4pt minus 4pt
     \bgroup
     \it
     \noindent
     \ignorespaces#1\unskip\ \egroup
     \ignorespaces}
\let\ts=\thinspace
\def\footnoterule{\kern-3pt\hrule width 2true cm\kern2.6pt}
\newcount\footcount \footcount=0
\def\advftncnt{\advance\footcount by1\global\footcount=\footcount}
\def\fonote#1{\advftncnt$^{\the\footcount}$\begingroup\petit
\parfillskip=0pt plus 1fil
\def\textindent##1{\hangindent0.5\oldparindent\noindent\hbox
to0.5\oldparindent{##1\hss}\ignorespaces}%
\vfootnote{$^{\the\footcount}$}{#1\vskip-9.69pt}\endgroup}
\def\item#1{\par\noindent
\hangindent6.5 mm\hangafter=0
\llap{#1\enspace}\ignorespaces}

\def\newenvironment#1#2#3#4{\long\def#1##1##2{\removelastskip
\vskip\baselineskip\noindent{#3#2\if>##1>.\else\unskip\ \ignorespaces
##1\unskip\fi\ }{#4\ignorespaces##2}\vskip\baselineskip}}
\newenvironment\lemma{Lemma}{\bf}{\it}
\newenvironment\proposition{Proposition}{\bf}{\it}
\newenvironment\theorem{Theorem}{\bf}{\it}
\newenvironment\corollary{Corollary}{\bf}{\it}
\newenvironment\example{Example}{\it}{\rm}
\newenvironment\exercise{Exercise}{\bf}{\rm}
\newenvironment\problem{Problem}{\bf}{\rm}
\newenvironment\solution{Solution}{\bf}{\rm}
\newenvironment\definition{Definition}{\bf}{\rm}
\newenvironment\note{Note}{\it}{\rm}
\newenvironment\question{Question}{\it}{\rm}
\long\def\remark#1{\removelastskip\vskip\baselineskip\noindent{\it
Remark.\ }\ignorespaces}
\long\def\proof#1{\removelastskip\vskip\baselineskip\noindent{\it
Proof\if>#1>\else\ \ignorespaces#1\fi.\ }\ignorespaces}
\def\typeset{\petit\noindent This article was processed by the author
using the \TeX\ macro package from Springer-Verlag.\par}
\outer\def\byebye{\bigskip\bigskip\typeset
\footcount=1\ifx\speciali\undefined\else
\loop\smallskip\noindent special character No\number\footcount:
\csname special\romannumeral\footcount\endcsname
\advance\footcount by 1\global\footcount=\footcount
\ifnum\footcount<11\repeat\fi
\vfill\supereject\end}

\def\12{{1\ov 2}}

\def\ov{\over}

\def\krig#1{\vbox{\ialign{\hfil##\hfil\crcr
$\raise0.3pt\hbox{$\scriptstyle \circ$}$\crcr\noalign
{\kern-0.02pt\nointerlineskip}
$\displaystyle{#1}$\crcr}}}

\contribution{Threshold Pion Photo- and Electroproduction in Chiral
Perturbation
Theory}
\author{V\'eronique Bernard}\footnote{}{\hskip -0.6truecm
Talk given at the workshop on Chiral Dynamics: Theory and Experiments,
MIT, Cambridge, USA,
 July 25 - July 29, 1994.}
\address{Centre de Recherches Nucl\'eaires et Universit\'e Louis Pasteur de
Strasbourg, Physique Th\'eorique, Bat. 40, BP28,
67037 Strasbourg Cedex, France}
\vskip -2truecm
\titlea{1}{Introduction}
Threshold pion photo- and electroproduction off nucleons allow us, by the
use of a well-understood probe, to
test our understanding of the strong interaction at low energies, i.e. in the
non--perturbative regime. Over the last few years, renewed interest in these
reactions has
emerged. This was first triggered through precise new data on neutral pion
photoproduction [1] which lead to a controversy about their theoretical
interpretations. Furthermore, new  data on
$\pi^0$ electroproduction [2] being obtained with an unprecedented accuracy
have given further constraints on the
understanding of these fundamental processes in the non--perturbative regime of
QCD.
While the study of these reactions, also with charged pions in the final
state, continues on the experimental as well as on the theoretical side,
complementary information can be gained from the two-pion production process
$\gamma N \to \pi \pi N$, where N denotes the nucleon and $\gamma$ the real
or virtual photon. Presently available data on this reaction focus on the
resonance region (excitation energies close to the first strong
resonance excitation
of the nucleon, the $\Delta$(1232))  and above, however, an
attempt to measure the $\pi^0 \pi^0 p$ channel very
close to threshold is actually being made in Mainz [3].

I will report here on a systematic analysis of the processes
$\gamma^{(\star)} N \to \pi N$ and $\gamma N \to \pi \pi N$ in the threshold
region making use of baryon chiral perturbation theory (CHPT). Other aspects
of nucleon chiral perturbation theory are discussed in these proceedings
by Ulf-G. Mei{\ss}ner. In general, CHPT allows one to systematically
investigate
the strictures of the spontaneously broken chiral symmetry of QCD. It is based
on the observation that in the three flavor sector of QCD, the quark masses are
small and that the theory in the limit of vanishing quark masses admits an
exact chiral symmetry. The latter is dynamically broken which leads to the
appearance of massless pseudoscalar excitations, the Goldstone bosons. In the
real word, the quark masses are not exactly zero and thus the Goldstone bosons
acquire a small mass. The interaction of these particles with each other
and matter fields like e.g. the nucleons are weak at low energies as mandated
by Goldstone's theorem. This fact is at the heart of CHPT which amounts to a
systematic and simultaneous expansion of the QCD Green functions in small
momenta and quark masses. To perturbatively restore unitarity it is mandatory
to consider pion loop diagrams. Here we will work in the one--loop
approximation
which has been shown to be of sufficient accuracy for many threshold phenomena.
For a review see H. Leutwyler [4] and Ulf-G. Mei{\ss}ner [5].

\titlea{2}{Effective Lagrangian}
In this section I will be rather short. For more details see A. Manohar and
Ulf-G. Mei{\ss}ner, these proceedings. To systematically work out the
consequences
of the spontaneously broken chiral symmetry at low energies, one makes use
of an effective Lagrangian of the asymptotically observed fields, in our case
the Goldstone bosons (pions) and the nucleons ({\it i.e.} proton and neutron).
We work in flavor SU(2) and mostly in the isospin limit $m_u = m_d = \hat m$
and consider terms up to and including ${\cal O}(q^4)$ where $q$ denotes
a generic small momentum,
in the heavy mass
formulation.
This formulation has the advantage that there is a one-to-one correspondence
between loops and powers of $q$. Indeed, this is not the case in the
relativistic
calculation where things are complicated by the fact that the nucleon mass
does not vanish in the chiral limit. The effective Lagrangian can be written
as:
$${\cal L}_{eff} = {\cal L}_{\pi N}^{(1)} + {\cal L}_{\pi N}^{(2)} +
 {\cal L}_{\pi N}^{(3)} +{\cal L}_{\pi N}^{(4)} + {\cal L}_{\pi \pi}^{(2,4)}
 \eqno(2.1)$$
where ${\cal L}_{\pi N}^{(1)}$ represents the non--linear $\sigma$ model
coupled
to nucleons. Four parameters enter the theory to lowest order, namely
the pion decay constant $F$, the axial coupling
constant $\krig g_A$, the nucleon mass $\krig m$ ($\krig m$ only enters
implicitely through
the non-relativistic energy--momentum relation $p= \krig m v$. It does not
appear
in the Lagrangian to lowest order)
 and the leading term in the quark mass expansion of
 the pion mass $M_{\pi}$.
$F$, $\krig g_A$ and $\krig m$ denotes quantities in the chiral limit, thus
differing from the physical ones through terms
proportional to the current quark mass.
The vertices  from ${\cal L}_{\pi N}^{(1)}$ needed to perform a one--loop
calculation of photo-  and electroproduction are
the pseudovector $\pi NN$ vertex and the well-known
Weinberg and Kroll Ruderman terms. Notice that working in the Coulomb
gauge $\epsilon_0 = 0$ (as we will do in the following)
there is no direct coupling of the
photon to the nucleon (proportional to $\epsilon \cdot v$). This of course
simplifies the calculation since many diagrams vanish.
${\cal L}_{\pi N}^{(2,3,4)}$ contain $1/m$ corrections and
counterterms. The a priori unknown coefficients of these counterterms
 are the so-called low energy constants. For a detailed
  discussion of their estimates
 see G. Ecker, these proceedings.
I will come back to
the ones entering the particular reactions discussed here
in section 3.1.
Note that ${\krig \kappa}_{S,V}$, the isoscalar and isovector
anomalous magnetic moments of the nucleon in the chiral limit, enter
${\cal L}_{\pi N}^{(2)}$ as such low energy constants.
Let me finally point out that another simplification arises
in the heavy mass formulation  from the fact that the algebra
only involves  the velocity operator $v$ and
the covariant spin operator $S$. For more details, see for example ref.[6].
\medskip
\titlea{3}{Threshold Pion Photoproduction}
Let us consider the reaction $\gamma (k) + N_1 (p_1) \to \pi^a (q) + N_2 (p_2)$
with $N_{1,2}$ denoting protons and/or neutrons and '$a$' refers to the charge
of the produced pion. In the case of real photons ($k^2 = 0$) one talks of
photoproduction whereas for virtual photons (radiated off an electron beam)
the process is called electroproduction. Of particular interest is the
threshold region where the photon has just enough energy to produce the pion at
rest or with a very small three-momentum. In this kinematical regime, it is
advantageous to perform a multipole decomposition since at threshold only the
$S$-waves survive. These multipoles are labelled $E, \, M, \, L_{l \pm}$, where
$E, \,
M, \, L$ stands for electric, magnetic, longitudinal respectively ($L$ of
course does
not appear in photoproduction where the photons are transverse),
 $l = 0,1,2, \ldots$ the pion orbital angular
momentum and
the $\pm$ refers to the total angular momentum of the pion-nucleon system, $j
=l \pm 1/2$. They parametrize the structure of the nucleon as probed with
low energy photons.
\titleb{3.1}{CHPT calculations}
Let us first concentrate on the photoproduction case. We will consider
here only the neutral channel. A detailed discussion of the
charged ones in the relativistic formalism is given in Ref.[7] and the
experimental status is reported by J. Bergstrom and M. Kovash
(these proceedings).
The T--matrix
depends on four amplitudes and takes the
following form close to threshold where it is legitimate to keep
only the $S$- and $P$-waves:
$$ T \cdot \vec \epsilon = \vec \sigma \cdot \vec \epsilon \, (E_{0+}
+ \hat k \cdot \hat q P_1)
+ \vec\sigma \cdot \hat k \, \vec \epsilon \cdot \hat q \,P_2
+ (\hat q \times \hat k ) \cdot \vec \epsilon \, P_3
\eqno(3.1) $$
The quantities $P_{1,2,3}$ represent the following combinations of the three
$P$-waves, $E_{1+}, \, M_{1+}$ and  $M_{1-}$,
$$\eqalign{
P_1 &= 3E_{1+} + M_{1+} - M_{1-}  \cr
P_2 &= 3E_{1+} - M_{1+} + M_{1-}  \cr
P_3 &= 2 M_{1+} + M_{1-} \cr} \eqno(3.2) $$
These four amplitudes are easily calculable within CHPT. Let us first discuss
the
${\cal O}(q^3)$ results. I should point out that since $\epsilon$, the
polarization vector of the photon,
is a quantity
of order $q$ in the chiral counting (it enters the covariant derivative at the
same level as $\partial_\mu $) $E_{0+}$ and the $P_i$ are therefore
quantities of order
$q^2$.
\item{--} At tree level one has contributions from ${\cal L}_{\pi N}
^{(2,3)}$ which bring corrections of order $E/m^2$ and $E^2/m^3$, respectively,
where $E$ is a generic small momentum which in the present case
can  either be $\omega$, the  $\pi^0$
center of mass energy,  $|\vec q|$ its three
momentum (for the $P$-waves) and $M_\pi$.
\item{--} Loops start at this order.
The well-known rescattering graph and the very important, as
we will see later, triangle
graph contribute to $E_{0+}$ while, as shown in Fig.1
two other graphs contribute to the $P$-waves.
It turns
out that these loop contributions are finite, of the type $E^2/F_\pi^3$
and  cancel for $P_3$.
One of
course should add the diagrams which contribute to mass and coupling constant
renormalization.
We don't show them here.
\item{--} Three counterterms appear at ${\cal O}(q^3)$. One, $\kappa_p$,
the anomalous magnetic moment of the proton in the chiral limit,
enters
${\cal L}_{\pi N}^{(2)}$ as discussed in the previous section.
Note that
in contrast to the relativistic result [8] $\kappa_p$ is almost entirely
given by the counterterm. It contributes to all the multipoles except $P_3$
at variance with the two other counterterms which only contribute to $P_3$.
These, $b_{22}$ and $b_{14}$ which come from ${\cal L}_{\pi N}^{(2)}$ and
${\cal L}_{\pi N}^{(3)}$ respectively,
have been first written down by Ecker [9]. They
are infinite but they enter through the combination $b =4b_{14} + g_A b_{22}$
which just cancels these infinities. Effectively there is thus
only one counterterm $b$
entering $P_3$.
There are two ways of determining these
counterterms (both will be investigated here). The first one which is in the
spirit of CHPT, is to fit them to the total and  differential
photocrossection. One then $\it predicts$ the multipoles and all the
electroproduction observables. An alternative way is to use the principle of
resonance saturation demonstrated by Ecker {\it et al.} [10] in the meson
sector.
The low energy constants are evaluated
through the exchange of resonances (in
our case the $\Delta, \rho$ and $\omega$). This is often called the QCD version
of Vector Meson Dominance. This introduces of course some scale dependence
since
the couterterms are obtained at a certain scale which is given by the resonance
masses. For more detail, see G. Ecker, these proceedings.
\medskip
\vskip 4truecm
\item{} Fig.1 One-loop diagrams contributing to $E_{0+}$ (two first)
 and $P_i$'s (two last).
\medskip
I will not enter into the discussion of the ${\cal O}(q^4)$ terms in detail
here.
Let me just say that the loops are calculated with one vertex from
${\cal L}_{\pi N}^{(2)}$ and one from ${\cal L}_{\pi N}^{(1)}$, or with
vertices
from ${\cal L}_{\pi N}^{(1)}$ but a propagator from ${\cal L}_{\pi N}^{(2)}$
or are
 $1/m$ corrections to the loops appearing at ${\cal O}(q^3)$. It turns out that
there are two more counterterms from ${\cal L}_{\pi N}^{(4)}$
contributing to $E_{0+}$ at this next order. Thus, the general expressions for
$E_{0+}$ and the $P$-waves to ${\cal O}(q^4)$ and ${\cal O}(q^3)$,
respectively,
are:
$$\eqalign{
E_{0+}&= {\alpha \over m^2 }E + \bigl({\beta \over m^3} +{\gamma \over F^3}
+\kappa_p \, \delta \bigr) E^2 + \bigl({\epsilon \over m^4} + {\mu \over m F^3}
\bigr) E^3 +d_5 M_\pi^2 \omega +d_6 \omega^3 \cr
P_{1,2}& = {\alpha '\over m^2} |\vec q \,| + \bigl({\beta ' \over m^3}
+{\gamma ' \over F^3} + \kappa_p \, \delta ' \bigr) |\vec q \,| E \cr
P_3& = {\rm small} \, \, {\rm Born} + d_7|\vec q \, | \omega \cr}
\eqno(3.3) $$
where $d_i$ are quantities proportional to the low energy constants,
$\alpha$, $\beta$, $\delta$, $\epsilon$ coming from tree diagrams are
real functions of $\omega$  and $\gamma$ and $\mu$ which are loop
contributions are complex function of $\omega$ for $\omega >
M_{\pi^+}$ (same holds for the ' quantities), thus restoring unitarity in
a perturbative manner.
An important fact is that
to lowest order $E_{0+}$ and the $P$'s have the same chiral power. Indeed,
$P_{1,2}$ are proportional to $| \vec q \,|$ and not $| \vec q \,| |\vec k \,|$
as it is usually assumed. However since $| \vec k \,|$ is a very slowly varying
function close to threshold it turns out that this approximation  is rather
good.
Note that apart from a small Born term $P_3$ is essentially given by a
counterterm.
\medskip
\titleb{3.2}{Low energy theorems}
A low energy theorem was derived in the seventies [11], giving $E_{0+}$
for neutral pion photoproduction at
threshold as a series in $\mu = M_\pi/m = 1/7$ up to and including ${\cal
O}(\mu^2)$ in terms of physical quantities only. This ``LET'' was based on
chiral symmetry and on some so--called harmless assumptions. It turned out
that it was in contradiction with the latest experimental informations [1].
The theorem
has been therefore reconsidered by several authors [12]
(and confirmed by some)
and the interpretation of the experimental data has been critically examined
[13,14].
Before entering the discussion of this LET, let me first emphasize that CHPT
is $not$ a model but is a method for solving QCD at low energy. As it is
well--known and most clearly spelled out by Weinberg [15] CHPT embodies
the very general principle of gauge invariance, analyticity, crossing and
PCAC (pion pole dominance). Therefore to lowest order, one recovers the
venerable current algebra LETs, which are based on these principles. The
effective chiral lagrangian is simply a tool to calculate these but also
all next-to-leading order corrections in a systematic and controlled fashion.
Thus any theory or model which claims to embody PCAC should lead to the same
result as CHPT if one goes to the same level of sophistication. Performing
a one-loop CHPT calculation, $E_{0+}$ at threshold is given by:
$$E_{0+} = -{e g_{\pi N} \over 8 \pi m} \mu \biggl\lbrace 1 - \bigl[ {1 \over
2} (3 + \kappa_p ) + ({m \over 4 F_\pi})^2 \bigr] \mu + {\cal O}(\mu^2 )
\biggr\rbrace \quad ,
\eqno(3.4)$$
with $\kappa_p = 3.71$ the anomalous magnetic moment of the proton.
The second term in the square brackets was not appearing in the original
``LET''.
It is the new one found in the
CHPT calculation.
It stems from the so-called triangle
diagram
and its crossed partner.
If one expresses $E_{0+}$ to lowest order in terms of
the
conventional Lorentz invariant function $A_1$, $E_{0+} = \mu A_1$,
one finds that
these diagrams give to leading order,
$\delta A_1 = ( e g_{\pi N} / 32
F_\pi^2 ) \, \mu$
due to the presence of a logarithmic
singularity and consequently {\it contributes
at next-to-leading order} in the quark mass
expansion of $E_{0+}$.
Thus this novel term originates from the presence of Goldstone bosons
which leads  to the existence of non analytical terms  ($\mu \sim \sqrt{\hat
m}$)
  invalidating the
assumptions ($\delta A_1 = {\cal O}(\mu^2)$)
used to derive what one should {\it not} call a ``LET'' but
a ``LEG'' (low energy
guess as defined by G. Ecker, this workshop).
Clearly, the expansion in $\mu$ is slowly converging, the
coefficient of the term of order $\mu^2$ is so large that it compensates
the leading term proportional to $\mu$. Therefore, for
a meaningful
prediction one has to go further in the expansion. To order $\mu^3$ one gets
$-1.14$
using the conventional units of $10^{-3} / M_{\pi^+}$ to be compared with
-3.35  at leading order and 0.89 at next-to-leading order. These results
show that the LET for $E_{0+}$ is practically useless and that it turns
out to be very hard to determine this quantity theoretically, making it not
the best test of chiral dynamics.

However, it is very easy to derive LETs for the $P$-waves within CHPT.
These have never been looked at before. They take the following forms:
$$\eqalign{ {1 \over |\vec q \,|}
P_1& = -{e g_{\pi N} \over 8 \pi m^2}  \biggl\lbrace 1 + \kappa_p
+ \bigl[
-1 - {\kappa_p \over 2} + {g_{\pi N}^2 \over 48 \pi}(10 - 3 \pi)
\bigr] \mu + {\cal O}(\mu^2 )
\biggr\rbrace \quad , \cr
 {1 \over |\vec q \,|}
P_2& = -{e g_{\pi N} \over 8 \pi m^2}  \biggl\lbrace - 1 - \kappa_p
+ \bigl[
3 + \kappa_p  - {g_{\pi N}^2 \over 12 \pi}
\bigr] {\mu \over 2} + {\cal O}(\mu^2 )
\biggr\rbrace \quad .     \cr}
\eqno(3.5)$$
They are very fastly converging functions of $\mu$. Indeed the terms of ${\cal
O}
(\mu)$ contributes for 6\% (less than 0.1\%) to $P_1$ ($P_2$) respectively.
They thus constitute a very  good test of chiral dynamics. Writing them in the
conventional  units $| \vec q \,| |\vec k \,| \, 10^{-3}/M_{\pi^+}$, one gets
10.3 and -10.9 for $P_1$ and $P_2$, respectively. Notice that already at next
to leading order $P_1$ and $P_2$ differ. Consequently, the value $E_{1+}\equiv
0$ is
excluded. I will end the discussion on the $P$-waves here just emphasizing
that they
depend weakly on energy, they show no cusp effect and they have a small
imaginary part.
\medskip
\titleb{3.3}{Results and discussion}
All the CHPT results presented in this section are preliminary.
Let me first concentrate on the real part of $E_{0+}$.
The reanalysis of the data agree on
its value at threshold,  $E_{0+} \sim -2.\,10^{-3}/M_{\pi^+}$
({\it accidentally} the experimental result and the LEG are in agreement)
but
disagree on its energy dependence, the difference between the
\medskip
\vskip 8truecm
\item{} Fig.2 Real part of $E_{0+}$.
\vskip 0.2truecm
\noindent reanalysis
coming essentially from different assumptions on the $P$-waves. We just saw
that $P_1$ and $P_2$ are constrained since they obey
 LETs, these were however not known
at the time of the reanalysis. Bergstrom [13] and Tiator [13] show a rather
strong energy dependence while Bernstein [13] has a weak one. At present this
weak energy dependence seems to be ruled out by the LETs which are violated in
Bernstein's reanalysis. In fig.2 is shown the CHPT result. There, some isospin
breaking effects have been taken into account by differentiating in the loops
the charge pion mass from the neutral one. The two curves represent the two
different ways of estimating the low energy constants discussed previously. The
full curve corresponds to the best fit to the differential and total
cross-sections, the band corresponding to a $\pm 1$ standard deviation.
The dotted curve corresponds to the determination via resonance exchange
where the values of the off-shell parameters of the $\Delta \gamma N$ vertex
and
the $\Delta$ propagator are again obtained
by a  best
fit to the data constrained so that these values lie within the error bars
of the theoretical estimates.
One notes that the value of $d_7$ is somewhat
 independent of the  method used.
The value of $E_{0+}$ at threshold lies between -1 to -1.5 somewhat smaller
in magnitude than
the experimental one.
It is very much constrained by the bell shape type of the
angular distribution of the differential cross-sections for a photon energy
larger than 150 MeV as seen in Fig.3 which shows how good the fits are.
The energy dependence is weaker than in the reanalysis of Bergstrom and Tiator,
their value of Re$E_{0+}$
at the $\pi^+ n$ threshold being close to zero to be compared with
the CHPT one of -0.5. Thus the difference of $-2.$ between the two
threshold cannot be obtained within a ${\cal O}(q^4)$ calculation. We had
already
seen that the $\mu$ expansion of $E_{0+}$ is very slowly converging, this of
course demands a two-loop calculation to test the validity of the results
presented here. Of course
more precise measurements are needed too.
\medskip
\vskip 5truecm
\item{} Fig.3 Angular distribution of $d \sigma / d\Omega$: $E_\gamma =151.4$
(left), $=153.7$ MeV (right).
\vskip 0.2truecm

The imaginary part of $E_{0+}$ is a very important quantity to look at and this
for the following reasons. First, its result is genuine since no counterterm
is involved in its calculation. Second, it is related to the change of its real
part through a dispersion relation. This can also be seen very easily by using
a simple but however realistic model where  $E_{0+}= a - b \sqrt{1-E_\gamma/
E_\gamma^{\,(+)}}$ with  $E_\gamma^{\,(+)}$ the photon energy at the $\pi^+ n$
threshold (a good fit is obtained with $a = 0.3$ and $b=5.7$ in the units
$10^{-3}/M_{\pi^+}$). Another point has been stressed by Bernstein [16].
The imaginary
part of $E_{0+}$ is related to the $\pi N$ scattering lengths through the
Fermi Watson theorem. Thus a measure of the imaginary part would lead to
another
way of determining these important quantities. Within CHPT to ${\cal O}(q^4)$,
${\rm Im} E_{0+}$ stays below $10^{-3}/M_{\pi^+}$ for
 $E_\gamma < 160$ MeV. It is
too small, as we have already seen, to account for the strong energy dependence
of Re$E_{0+}$. For all these reasons it would be very interesting to have
a measure of  ${\rm Im} E_{0+}$. This can be done through a measurement of
polarized
observables. Two of them turn out to be particularly sensitive to this
quantity, these are
the polarized target asymmetry T and
the recoil polarization P as can easily be seen by looking at their
definitions:
$$\eqalign{
T & \propto {\rm Im}((E_{0+} + \cos \theta P_1)(P_2 - P_3) \cr
P & \propto -{\rm Im}((E_{0+} + \cos \theta P_1)(P_2 + P_3)
 \cr} \eqno(3.6) $$
The CHPT results are shown in Fig. 4. One sees that though P stays
very small, T reaches 15 to 20\% over a wide range of angles. This makes this
observable a good candidate for a measurement of ${\rm Im} E_{0+}$.
\medskip
\vskip 5.0truecm
\item{} Fig.4 Polarized observables $P$ and $T$ for $E_\gamma = 151.4$ MeV.
\medskip
\titlea{4.}{Threshold Pion Electroproduction}
\medskip
\vskip 8.5truecm
\item{} Fig.5 Angular distribution of $ d\sigma_T/d \Omega$ for $k^2 = -0.1$
GeV$^{-2}$.
\vskip 0.2truecm
I will now discuss some rather new data on electroproduction.
 Welch et al.[2] have published the
$S$-wave cross section for the reaction $\gamma^\star p \to \pi^0 p$
very close to the photon point. This measurement is a quantum step
compared to previous determinations
which mostly date back to the
seventies when pion electroproduction was still a hot topic in particle
physics. In this experiment, $k^2$ varied between -0.04 and -0.10
GeV$^{-2}$ and the $S$-wave cross section could be extracted with
an unprecedented accuracy
(see fig.2 in [2]). This is also the
kinematical regime where a CHPT calculation might offer some insight.
Indeed, in ref.[17] it was shown that
the $k^2$-dependence of this
cross section seems to indicate the necessity of loop effects. With
conventional models including e.g. form factors and the anomalous
magnetic moment coupling the trend of the data can not be described.
However, the corrections from the one loop diagrams to the tree level
prediction are substantial. This gives further credit to the previously
made
statement that a calculation beyond next-to-leading order should be
performed. Very recently, Distler {\it et al.} [3] have measured the angular
distribution of the differential transverse cross section at $k^2 = -0.1$
GeV$^{-2}$. Even though this value is somewhat too high for a good test of
CHPT (
there, the one-loop order corrections are of the order of 50\%) it is
interesting to see that the preliminary experimental results show the same
trends as the CHPT ones and are at variance with the pseudo-vector Born
approximation (see Fig.5). It would be of particular interest to have
experimental informations at $k^2 = -0.05$ GeV$^{-2}$.

The last topic I want to address in this section concerns
the determination of the nucleon  axial radius from charged pion
electroproduction. Let me briefly explain how the axial form factor
comes into play. The basic matrix element to be considered is the
time-ordered product of the electromagnetic (vector)  current with
the interpolating pion field sandwiched between nucleon states. Now
one can use the PCAC relation and express the pion field in terms of
the divergence of the axial current. Thus, a commutator of the form
[V,A] arises. Current algebra tells us that
this gives an axial current between the incoming and outgoing
nucleon fields and, alas, the axial form factor. The isospin factors
combine in a way that they form a totally antisymmetric combination
which can not be probed in neutral pion production. These ideas were
formalized in the venerable low-energy theorem (LET)
due to Nambu, Luri{\'e}
and Shrauner [18] for the isospin--odd electric dipole amplitude
$E_{0+}^{(-)}$ in the chiral limit,
$$
E_{0+}^{(-)}(M_\pi=0, k^2) ={e g_A \over 8 \pi
F_\pi} \biggl\lbrace 1 +{k^2 \over 6} r_A^2 + { k^2 \over 4m^2} (\kappa_V
+ {1 \over 2}) + {\cal O} (k^3) \biggr\rbrace
\eqno(2.15)$$
Therefore, measuring the reactions
$\gamma^\star p \to \pi^+ n$ and
$\gamma^\star n \to \pi^- p$ allows to extract
$E_{0+}^{(-)}$ and one can determine the axial radius of the nucleon, $r_A$.
This quantity measures the distribution of spin and isospin in the nucleon,
i.e. probes the Gamov--Teller operator $\bf  \sigma \cdot  \tau$. A
priori, the axial radius is expected to be different from the typical
electromagnetic size,
$r_{{\rm em}} \simeq 0.85$ fm. It is customary to parametrize
the axial form factor $G_A (k^2)$ by a dipole form, $G_A (k^2)/g_A = (1 - k^2 /
M_A^2 )^{-2}$ which leads to the relation $r_A = \sqrt{12} / M_A$. The axial
radius determined from electroproduction data is typically $r_A = 0.59 \pm
0.04$ fm
whereas (anti)neutrino-nucleon reactions lead to somewhat
larger values, $r_A = 0.65 \pm 0.03$ fm. This discrepancy is usually not
taken seriously since the values overlap within the error bars. However, it
was shown in ref.[19] that pion loops modify the LET (2.15) at order $k^2$
for finite pion mass. In the heavy mass formalism, the coefficient of the
$k^2$ term reads
$$ {1 \over 6} r_A^2 + {1 \over 4 m^2}(\kappa_V +{1 \over2}) +
{1 \over 128 F_\pi^2} (1 - {12 \over \pi^2})
\eqno(2.16)$$
where the last term in (2.16) is the new one. This means that previously one
had extracted a modified radius, the correction being $3 (1 - 12/\pi^2 ) / 64
F_\pi^2 \simeq -0.046$ fm$^2$. This closes the gap between the values of
$r_A$ extracted from electroproduction and neutrino data. It remains to be
seen how the $1 / m$ suppressed terms will modify the result (2.16). Such
investigations are underway.

\medskip
\titlea{5.}{Threshold Two Pion Photoproduction:}
At threshold in the center-of-mass frame ({\it i.e.} $\vec q_1 =\vec q_2=0$),
the two-pion photoproduction current matrix element can be decomposed
into amplitudes as follows if we work to first order in the
electromagnetic coupling $e$,
$$T \cdot \epsilon = \chi^\dagger _f \lbrace i \vec \sigma \cdot
 (\vec \epsilon \times \vec k) [M_1 \delta^{ab} +M_2 \delta^{ab} \tau^3
 +M_3 (\delta^{a3} \tau^b +\delta^{b3} \tau^a)]\rbrace \chi_i
 \eqno(4.1)$$
with $\chi_{i,f}$ two-component Pauli-spinors and isospinors and we used the
gauge $\epsilon_0 =0$. To leading order in the chiral expansion, only the
amplitudes $M_2$ and $M_3$ are non vanishing, with $M_2=-2M_3$ [20]. Therefore
the
production of two neutral pions is strictly suppressed.
One can derive LETs for $M_{2,3}$. They read:
$$M_2 = -2M_3 = {e \over 4 m F_\pi^2}(2 g_A^2 - 1 - \kappa_V) \eqno(4.2) $$
There exists in the literature some incomplete determination of these LETs
[21]. The authors did not use the most general effective Lagrangian and thus
did not get the term proportional to $\kappa_V$. At next order in the chiral
expansion, one has to consider kinematical ($1/m$) corrections, one-loop
contributions and tree graphs with insertion of the $\Delta(1232)$ (chirally
expanded). Note that, to this order, the $\Delta$ contributions are absent
in the $\pi^0 \pi^0$ channel.
Here we use the principle of resonance saturation to estimate the
counterterms. At this order one has a LET for $M_1$:
$$M_1 = {e g_A^2 M_\pi \over 4 m F_\pi^2} \eqno(4.2)$$
and the $M_{2,3}$ have a somewhat more complicated expressions (see ref.[20]).
Their loop contributions have a non-zero imaginary part even at threshold which
comes from the rescattering graph. Due to unitarity the pertinent loop
functions have a right hand cut starting at $s = (m +M_\pi)^2$ (the single
pion production threshold) and these functions are here to be evaluated at
$s = (m +2M_\pi)^2$ (the two- pion production threshold).In Fig.6 are shown
the total cross sections for $\gamma p \to \pi^+ \pi^- p$, $\pi^+ \pi^0 n$
and  $\pi^0 \pi^0 p$. Here the calculation was done with the correct
phase-space and approximating the amplitudes in the threshold region
through their threshold values in the case of the two last reactions. For the
first one the first correction above threshold is also shown [20].  At
$E_\gamma = 320$ MeV, the total cross section for $\pi^0 \pi^0$ production is
0.5 nb whereas the competing $\pi^+ \pi^0 n$ final state has $\sigma_{\rm tot}
= 0.07$ nb. Double neutral pion production reaches $\sigma_{\rm tot}= 1.0$ nb
at $E_\gamma = 324.3$ MeV in comparison to  $\sigma_{\rm tot}
(\gamma p \to \pi^0 \pi^+ n) = 0.26$ nb and $\sigma_{\rm tot}
(\gamma p \to \pi^+ \pi^- p) < 0.1$ nb. This means that for the first
$10 \ldots 12$ MeV above $\pi^0 \pi^0$ threshold (chiral window),
one has a fairly clean
signal and much more neutrals than expected. Of course, the above threshold
correction for all the channels should be calculated systematically. However
the first correction, which vanishes proportional to $|\vec q_i|$ (i=1,2)
at threshold, has been calculated and found to be small. It is therefore
conceivable that the qualitative features described above will not change
if even higher corrections are taken into account. We hope to report on
these in the not too distant future. From the experimental side TAPS seems to
have seen some $\pi^0$'s close to threshold. For more detail, see T. H.
Walcher,
these proceedings.

\medskip
\vskip 6.3truecm
\item{} Fig.6 Total cross sections (in nb) for the $\gamma p$ initial state
\medskip
\titlea{6}{Summary and Outlook}
Pion photo- and electroproduction in the threshold  region
is a good testing ground
of the chiral nucleon structure. A one-loop CHPT calculation gives
a satisfactory description of most of the existing observables. The pertinent
results of our investigation can be summarized as follows:
\item{-}The LET for $E_{0+}$ is practically useless. At variance with what
is usually believed the $P$-waves are the quantities providing a good test
of chiral dynamics.
\item{-}We have stressed the importance of the imaginary part of $E_{0+}$.
\item{-}We have obtained a better understanding of the axial radius of the
nucleon as measured in charged electroproduction.
\item{-}We have seen that there is a window of about 10 MeV
above $\pi^0 \pi^0$ threshold in which one should detect much more neutrals
than expected in two pion photoproduction.

At this stage some more efforts are needed on the theoretical side to pin
down the isospin breaking effects. Also a
 two loop calculation is mandatory to
determine the validity of the results presented here, especially in the case
of the S-wave observables. On the experimental side, it is clear that in order
to test CHPT, one has to measure close to threshold and to get to very accurate
data. I want to stress again that it would be very important to have a measure
of the imaginary part of $E_{0+}$ as well as a good determination of the energy
dependence of its real part.
\medskip
\titlea{7}{Acknowledgements}

I would like to thank the organizers
for
their invitation and kind hospitality.
\vskip 0.05truecm
\begref{References}{[MT1]}
\refno{[1]}
E. Mazzucato et al., {\it Phys. Rev. Lett.\/}
{\bf 57} (1986) 3144;
R. Beck et al., {\it Phys. Rev. Lett.\/}
{\bf 65} (1990) 1841
\refno{[2]}
T. P. Welch et al., {\it Phys. Rev. Lett.\/}
{\bf 69} (1992) 2761
\refno{[3]}
T. H.  Walcher, these proceedings
\refno{[4]}
H. Leutwyler, see for exemple these proceedings and ``Principles of Chiral
Perturbation Theory'', lectures given at the Workshop ``Hadrons 1994'',
Gramado, Brasil, BUTP-94/13
\refno{[5]}
Ulf-G. Mei{\ss}ner,
{\it Rep. Prog. Phys.\/}
{\bf 56} (1993) 903
\refno{[6]}
V. Bernard, N. Kaiser and Ulf-G. Mei{\ss}ner,
{\it Z. Phys.\/}
{\bf A348} (1994) 317; in preparation
\refno{[7]}
V. Bernard, N. Kaiser and Ulf-G. Mei{\ss}ner,
{\it Nucl. Phys.\/}
{\bf B383} (1992) 442
\refno{[8]}
J. Gasser, M.E. Sainio and A. ${\check {\rm S}}$varc,
{\it Nucl. Phys.\/}
 {\bf B307} (1988) 779
\refno{[9]}
G. Ecker,
``Chiral Invariant Renormalization of the Pion-Nucleon Interaction'',
{\it Phys. Lett.\/}
{\bf B}, in print, UWThPh-1994-1
\refno{[10]}
G. Ecker, J. Gasser, A. Pich and E. de Rafael,
{\it Nucl. Phys.\/}
 {\bf B321} (1989) 311;
G. Ecker, J. Gasser, H. Leutwyler, A. Pich and E. de Rafael,
{\it Phys. Lett.\/}
{\bf B223} (1989) 425
\refno{[11]}
A. I. Vainshtein and V. I. Zakharov,
{\it Nucl. Phys.\/}
 {\bf B36} (1972) 589; P. de Baenst,
{\it Nucl. Phys.\/}
 {\bf B24} (1970) 633
\refno{[12]}
V. Bernard, N. Kaiser, J. Gasser and Ulf-G. Mei{\ss}ner,
{\it Phys. Lett.\/}
{\bf B268} (1991) 291 and references therein
\refno{[13]}
A. M. Bernstein and B. R. Holstein,
{\it Comments Nucl. Par. Phys.\/}
{\bf 20} (1991) 197;
J. Bergstrom,
{\it Phys. Rev.\/}
{\bf C44} (1991) 1768;
L. Tiator,
''Meson Photo- and Electroproduction'', lecture given at the II TAPS
Workshop, Alicante, 1993
\refno{[14]}
R. Davidson and N. C. Mukhopadhyay,
{\it Phys. Rev. Lett.\/}
{\bf 60} (1988) 748;
E. Mazzucato et al.,
{\it Phys. Rev. Lett.\/}
{\bf 60} (1988) 749;
S. Nozawa, T.-S.H. Lee and B. Blankleider,
{\it Phys. Rev.\/}
{\bf C41} (1990) 213;
A. N. Kamal,
{\it Phys. Rev. Lett.\/}
{\bf 63} (1989) 213
\refno{[15]}
S. Weinberg,
{\it Physica\/}
{\bf A96} (1979) 327
\refno{[16]}
A. M. Bernstein,
$\pi N$ {\it Newsletter\/}
 {\bf 9} (1993) 55
\refno{[17]}
V. Bernard, N. Kaiser, T.-S.H. Lee and Ulf-G. Mei{\ss}ner,
{\it Phys. Rev. Lett.} {\bf 70} (1993) 387;
{\it Phys. Reports\/}, in print
\refno{[18]}
Y. Nambu and D. Luri\'e, {\it Phys. Rev.\/} {\bf 125}
(1962) 1429;
Y. Nambu and E. Shrauner, {\it Phys. Rev.\/} {\bf 128}
(1962) 862
\refno{[19]}
V. Bernard, N. Kaiser and Ulf-G. Mei{\ss}ner,
{\it Phys. Rev. Lett.\/} {\bf 69} (1992) 1877
\refno{[20]}
V. Bernard, N. Kaiser and Ulf-G. Mei{\ss}ner,
``Threshold two-pion photo- and electroproduction: More neutrals than
expected'', CRN-94/14, {\it Nucl. Phys.\/} {\bf A}, in print
\refno{[21]}
R. Dahm and D. Drechsel, in Proc. Seventh Amsterdam Mini-Conference, eds.
H. P. Blok, J. H. Koch and H. De Vries, Amsterdam, 1991;
M. Benmerrouche and E. Tomusiak,
{\it Phys. Rev. Lett.\/} {\bf 73} (1994) 400
\endref
\byebye

